\documentclass[twocolumn,showpacs,preprintnumbers,amsmath,amssymb,floatfix,prc]{revtex4}
\usepackage{graphicx}
\usepackage{dcolumn}
\usepackage{bm}
\def\htm{\hat{t}}

\def\beq{\begin{equation}}
\def\eeq{\end{equation}}
\def\be{\begin{eqnarray}}
\def\ee{\end{eqnarray}}

\newcommand{\f}[2]{\frac{#1}{#2}}
\newcommand{\dd}  { {\textrm d}}

\newcommand{\lsim}{
 \mathrel{\setbox0=\hbox{$<$}\raise0.6ex\copy0\kern-\wd0
 \lower0.65ex\hbox{$\sim$}}}

\newcommand{\gsim}{
 \mathrel{\setbox0=\hbox{$>$}\raise0.6ex\copy0\kern-\wd0
 \lower0.65ex\hbox{$\sim$}}}

\begin{document}
\title{Hadron production in deuteron-gold collisions and nuclear parton 
distributions}
\author{Adeola Adeluyi}
\author{George Fai}
\affiliation{Center for Nuclear Research, Department of Physics \\
Kent State University, Kent, OH 44242, USA}

\date{\today}
\begin{abstract}
We calculate nuclear modification factors $R_{dAu}$, 
central-to-peripheral ratios, $R_{CP}$, and
pseudorapidity asymmetries $Y_{Asym}$ in deuteron-gold collisions at 
$\sqrt{s} = 200$ GeV in the framework of leading-order (LO) perturbative 
Quantum Chromodynamics. We use the Eskola-Kolhinen-Salgado (EKS),
the Frankfurt-Guzey-Strikman (FGS) and the Hirai-Kumano-Nagai (HKN) 
nuclear parton distribution functions and the Albino-Kramer-Kniehl 
(AKK) fragmentation functions in our calculations. Results are 
compared to experimental data from the BRAHMS and STAR collaborations.
\bf\end{abstract}
\maketitle
\vspace{1cm}
%
%
\section{Introduction}
\label{intro}

In the rapidly-developing field of relativistic nuclear collision physics,
questions related to the distribution of partons in nucleons and nuclei are
of great current interest. The modification of the parton distribution 
functions (PDFs) in the nuclear environment has attracted growing attention 
ever since the pioneering EMC experiment\cite{Aubert:1983xm,Geesaman:1995yd}.
At the Relativistic Heavy Ion Collider (RHIC), we are well-positioned 
to study the nuclear PDFs (nPDFs) in a center-of-mass energy and 
transverse-momentum domain where perturbative Quantum Chromodynamics (pQCD) 
is expected to work well. Thus, hadron production in nuclear collision experiments 
at RHIC should provide information on how such nonperturbative ingredients of
pQCD as the parton distribution functions are modified by the presence of 
the nuclear medium.
 
However, the description of nuclear collisions based on pQCD is a complicated 
task. Much of the complication derives from the intrinsically complex nature 
of the nuclear environment in collisions of heavy nuclei. In addition to the 
initial-state modification of the PDFs, the cross section of high-$p_T$ 
hadron production is influenced by final-state effects (such as jet 
energy loss) and a complicated geometry. To better understand the physics of 
pQCD in the nuclear environment, it is highly desirable to disentangle the 
different nuclear phenomena affecting high-$p_T$ ($p_T \gsim 2$ GeV/c) hadron 
production. These phenomena (not present in proton-proton collisions) manifest 
themselves in e.g. the measured nuclear modification factors.  

Deuteron-gold (d+Au) collisions provide a good compromise and testing ground 
for these purposes.  The deuteron, being the simplest ``real'' nucleus, affords 
a complexity higher than a proton, but much less than that of a typical heavy 
nucleus like gold. Thus, a good understanding of d+Au collisions is invaluable 
in elucidating the added complexity associated with collisions of more complex 
heavy nuclei. Accordingly, d+Au collisions have been extensively studied at
RHIC (see e.g. \cite{PHENIXdAu,STARdAu,Arsene:2004ux,Yang:2007qr}). A new feature,
offered by nonidentical colliding beams like d+Au, is the pseudorapidity asymmetry,
examined in some detail recently by the STAR collaboration\cite{Abelev:2006pp}. 

Any pQCD calculation involves, in addition to partonic differential cross sections,
parton distribution functions (PDFs) and fragmentation functions (FFs) to connect 
to the observable level. The latter ingredients are non-perturbative, but universal
in the absence of the nuclear environment. Typical pQCD-based calculations of 
nucleus-nucleus collisions use modified (nuclear) PDFs and deal with issues like
jet energy loss (jet quenching) and collision geometry. In this work, we focus 
on the PDFs (and of course can not avoid treating the geometry of the collision). 
Even in the proton (nucleon), our knowledge of the PDFs is naturally limited; 
the nPDFs are much less well known in the wide range of momentum fraction~$x$  
needed for reliable calculations. The nuclear gluon distribution, in particular, 
is poorly constrained. The uncertainties in the nPDFs directly affect the accuracy 
of pQCD calculations. It is therefore important not to rely on a single nPDF 
parametrization. Calculations utilizing different parameterizations offer a useful 
check on the performances of the different nPDFs in describing relevant observables.

In the present work we compute nuclear modifications expressed in terms of the 
ratios $R_{dAu}$ (nuclear modification factor, see eq.~(\ref{RdAu})) and $R_{CP}$
(central-to-peripheral ratio, see eq.~(\ref{RCP})). We also calculate the 
pseudorapidity asymmetry $Y_{Asym}$ (see eq.~(\ref{yasym})) in certain pseudorapidity 
intervals. We focus attention on the phenomenon of nuclear shadowing, the difference 
between PDFs and nPDFs, leaving aside possible additional effects like jet quenching 
or intrinsic parton transverse momentum and its broadening in nuclear collisions. 
In this way we establish a minimalist base line for experimental comparisons.

An earlier study along these lines~\cite{Vogt:2004hf} used the 
Eskola-Kolhinen-Salgado (EKS)\cite{Eskola:1998df} and 
Frankfurt-Guzey-Strikman (FGS)\cite{Frankfurt:2003zd} parameterizations 
of nuclear shadowing. In Refs.~\cite{Eskola:1998df,Frankfurt:2003zd}
the basic object is the shadowing function, which encodes the relevant nuclear
information. The nPDFs are then a product of the shadowing function 
and nucleon PDFs. In our present study we are particularly interested in
the Hirai-Kumano-Nagai (HKN) parameterization~\cite{Shad_HKN}, which
also gives the parton distribution functions for the deuteron, unlike 
other parameterizations in which the deuteron is not shadowed.  (Deuteron 
shadowing is of course a small effect compared to shadowing in heavy nuclei.)
A short review of the available nuclear parton distribution functions and
their differences can be found in Ref.~\cite{Kolhinen:2005az}. Another 
point of departure in the present study from Ref.~\cite{Vogt:2004hf} is 
the use here of the Albino-Kramer-Kniehl (AKK) fragmentation 
functions~\cite{Albino:2005me}. These updated fragmentation functions 
are only available in next-to-leading order (NLO) and incorporate new 
experimental information from the OPAL Collaboration, including light quark 
tagging probabilities\cite{Abbiendi:1999ry}. 
Thus AKK is expected to offer a better description of the fragmentation process. 
For calculations involving the EKS and FGS shadowing functions we need the
nucleon parton distributions. We use the MRST2001 leading order (LO)
PDFs\cite{Martin:2001es} for consistency with HKN, where the underlying 
nucleon PDFs are the MRST2001 LO PDFs. 

The paper is organized as follows: in Sec.~\ref{formalism} we review the basic 
formalism of LO pQCD as applied to d+Au collisions. This Section also includes
the definitions of the nuclear modification factor, the central-to-peripheral
ratio, and the pseudorapidity asymmetry, and a brief review of the available 
experimental data. We present the results of our calculation in Sec.~\ref{res}, 
and conclude in Sec.~\ref{concl}.

\section{Basic Formalism}
\label{formalism}

The invariant cross section for the d+Au $\!\to\!$ h+X reaction, 
with respect to pseudorapidity $\eta$ and transverse momentum $p_T$ 
can be written as
\begin{eqnarray} \label{eq:dAu}
\f{\dd\sigma_{dAu}^{h}}{\dd\eta d^2p_T} =   
\sum_{\!\!abcd}\! \int\!\!\dd^2b \ \dd^2s \ \dd x_a \dd x_b \dd z_c \   
t_{d}({\vec s}) \ t_{Au}({|\vec b- \vec s|}) \nonumber \\
F_{\!a/d}(x_a,Q^2,\!{\vec s},z)\
F_{\!b/Au}(x_b,Q^2,\!{|\vec b- \vec s|},z')\ \nonumber \\
\f{\dd\sigma(ab\!\to\!cd)}{\dd\htm}\,
\f{D_{h/c}(z_c,\! Q_{f}^2)}{\pi z_c^2} \hat{s} \, \delta(\hat{s}+\htm+\hat{u}) \,\, ,
\end{eqnarray}
where $x_a$ and $x_b$ are parton momentum fractions in deuteron and gold, respectively, 
and $z_c$ is the fraction of the parton momentum carried by the final-state hadron~$h$.
The factorization and fragmentation scales are $Q$ and $Q_f$, respectively. 
Here,
\begin{equation} \label{eq:tf}
t_{A}({\vec s}) = \int\!\! \dd z \rho_{A}(\!{\vec s},z)
\end{equation}
is the Glauber thickness function of nucleus $A$, with the nuclear density distribution,
$\rho_{A}({\vec s},z)$ subject to the normalization condition
\begin{equation}
\int\!\! \dd^2s \ \dd z \rho_{A}({\vec s},z) = A \,\, .
\end{equation}
The quantity $\dd\sigma(ab\!\to\!cd)/\dd\htm$ in eq.~(\ref{eq:dAu}) represents the 
perturbatively calculable partonic cross section, and $D_{h/c}(z_c,\!{Q_{f}}^2)$ stands 
for the fragmentation function of parton $c$ to produce hadron $h$, evaluated at momentum 
fraction $z_c$ and fragmentation scale $Q_f$. 
Using the $\delta$-function in eq. (\ref{eq:dAu}), the integration over $ z_c$ can
be carried out explicitly. Integration limits over $x_a$ and $x_b$ are then
($x_{amin},1$) and ($x_{bmin}(x_a),1$) respectively. Note that $x_{bmin}$ is
a function of $x_a$. In addition, $z_c$ is also a function of both $x_a$ and $x_b$.

In the present study, we are primarily concerned with $F_{\!a/A}(x,Q^2,\!{\vec s},z)$, 
the nuclear parton distribution function (nPDF) for nucleus $A$. In light of the nuclear
modifications discussed in Sec.~\ref{intro}, it is natural to assume that the nPDF
depends on the location of the parton in the nucleus, (${\vec s},z)$ (or at least on
its position relative to the beam axis $\vec s$). To connect this ``inhomogeneous''
nPDF to the geometry-independent (homogeneous) nPDF ${\cal F}_{\!a/A}(x,Q^2)$, the 
normalization condition
\beq
\int\!\! \dd^2s \ \dd z \,
\rho_A({\vec s},z) F_{\!a/A}(x,Q^2,\!{\vec s},z) = {\cal F}_{\!a/A}(x,Q^2)
\eeq 
should be satisfied. 

In the EKS\cite{Eskola:1998df},
FGS\cite{Frankfurt:2003zd}, and HIJING\cite{Li:2001xa} 
parameterizations the (homogeneous) shadowing function 
${\cal S}(x,Q^2)$ is introduced, and the nPDF is written as
\beq
{\cal F}_{\!a/A}(x,Q^2) = {\cal S}(x,Q^2) f_{\!a/N}(x,Q^2) \,\, ,
\eeq
%
where 
$f_{\!a/N}(x,Q^2)$ is the PDF of the nucleon, which can be expressed as 
\begin{equation}
f_{\!a/N}(x,Q^2) = 
\f{Z}{A} f_{a/p}(x,Q^2) + (1-\f{Z}{A}) f_{a/n}(x,Q^2) \,\, ,
\end{equation}
with $f_{a/p}(x,Q^2)$ [$f_{a/n}(x,Q^2)$] being the proton [neutron] parton
distribution function as a function of Bjorken $x$ and factorization scale
$Q$. The HKN parameterization is already given in terms of nPDFs.

Especially at small values of $x$, where coherence effects are important, 
scaling with the thickness function appears to be more physical than
with the local density. Therefore, in this work we assume that shadowing 
is proportional to the thickness function (\ref{eq:tf})  
(i.e. we adopt the second option discussed in Ref.~\cite{Vogt:2004hf}).

The connection between the inhomogeneous and homogeneous shadowing functions,
$S_A(x,Q^2,\!{\vec s},z)$ and ${\cal S}(x,Q^2)$, respectively,
can be written as 
\beq
S_A(x,Q^2,\!{\vec s},z) = 1 + N [{\cal S}(x,Q^2) - 1 ] 
\f{\int\!\! \dd z \rho_{A}(\!{\vec s},z)}{\int\!\! \dd z \rho_{A}(0,z)} \,\, ,
\eeq
with $N$ a normalization constant. In other words, the deviation of the 
inhomogeneous shadowing function from unity is proportional to the deviation
from unity of the homogeneous shadowing function and proportional to the
thickness function. While inhomogeneous FGS nPDFs are  
available, for consistency and ease of comparison with EKS and HKN,  
we apply the homogeneous FGS nPDFs.

We obtain the density distribution of the deuteron from the Hulthen 
wave function\cite{Hulthen1957} (as in Ref.~\cite{Kharzeev:2002ei}), while a 
Woods-Saxon density distribution is used for gold with parameters from
Ref.~\cite{DeJager:1974dg}. Since the Nijmegen deuteron wave function\cite{Nijm},
which we also applied, gives similar results to the Hulthen wave function, 
we report only the calculations using the Hulthen wave function here.

We fix the scales as $Q = Q_f = p_T$, where $p_T$ is the final 
hadronic transverse momentum. We also carried out calculations with
the scales $Q = p_T/z_c$, $Q_f = p_T$. Results with the latter choice do
not differ significantly from those obtained by having both scales fixed at $p_T$.
The partonic differential cross sections, $\dd\sigma(ab\!\to\!cd)/\dd\htm$ 
were evaluated at leading order (LO). We note that, if a $K$ factor was used to  
approximate the effects of higher orders, these effects would cancel
in the ratios calculated in the present study. For the fragmentation functions
we use the AKK set\cite{Albino:2005me} throughout.
%
\subsection{Nuclear Modification Factors}
\label{modfacts}
The d+Au nuclear modification factor, $R_{dAu}$ is defined as 
\begin{equation}
\!R_{dAu}(p_{T}) =
 \f{1}{\langle N_{bin} \rangle} \, \f{\dd\sigma_{dAu}^{h}}{\dd\eta \ \dd^2p_T} \left/
 \f{\dd\sigma_{pp}^{h}}{\dd\eta \ d^2p_T} \right. \,\, ,
\label{RdAu}
\end{equation}
where the average number of binary collisions, $\langle N_{bin} \rangle$ 
in the various impact-parameter bins is given by
\begin{equation}
  \langle N_{bin} \rangle = \langle \sigma_{NN}^{in}\ T_{dAu}(b) \rangle \,\, .
\end{equation}
Here $\sigma_{NN}^{in}$ is the inelastic nucleon-nucleon cross section, and 
\beq
T_{dAu}(b) = \int\!\! \dd^2s \ t_{d}({\vec s}) \ t_{Au}({|\vec b- \vec s|})
\eeq
represents the deuteron-gold nuclear overlap function. The nuclear modification 
factor $R_{dAu}$ is thus just the ratio of the d+Au and proton-proton (pp) cross 
sections, normalized by the average
number of binary collisions,~$\langle N_{bin} \rangle$. 

\subsection{Central-to-Peripheral Ratios}
\label{CPratios}

A related ratio, which dispenses with the need for a reference 
pp cross section and uses information from the same experiment in 
numerator and denominator, thus canceling most systematic errors,
is the central to peripheral ratio defined as
\begin{equation}
\!R_{CP}(p_{T}) =
    \f{1}{\langle N_{bin} \rangle_{C}}
         \f{\dd\sigma_{dAu}^{h \, C}}{\dd\eta \ \dd^2p_T} \left/    
    \f{1}{\langle N_{bin} \rangle_{P}}
   \f{\dd\sigma_{dAu}^{h \, P}}{\dd\eta \ \dd^2p_T}       \right.  \,\, ,
\label{RCP}
\end{equation}
where $\langle N_{bin} \rangle$ is as defined above. The label $C$ stands
for the central event class, while $P$ denotes the peripheral class. The
centrality classes are chosen according to centrality cuts on the
experimental data.

The nuclear modification factors $R_{dAu}$ have been measured at several
pseudorapidities by the BRAHMS Collaboration, and are presented at
$|\eta| \le 0.2 \, (\eta = 0)$, $0.8 \le \eta \le 1.2 \, (\eta = 1)$, 
$1.9 \le \eta \le 2.35 \, (\eta = 2.2)$, and 
$2.9 \le \eta \le 3.5 \, (\eta=3.2)$\cite{Arsene:2004ux}.
At small rapidities ($\eta = 0,1$), the data are given for the sum of 
charged hadrons, while at forward rapidities negatively charged 
hadron data are available. The AKK fragmentation functions are for 
average charged hadrons. It should be remembered that the calculated results 
for all rapidities are therefore for the average of charged hadrons.
Furthermore, the BRAHMS data are given in 
three centrality classes: central (0-20)\%, semicentral (30-50)\%, and 
peripheral (60-80)\% (as a percentile of the geometric cross section). 
With a Woods-Saxon density for gold and the Hulthen wave function for the deuteron, 
a Glauber calculation of $T_{dAu}$ relates these classes to impact parameter
intervals as\cite{Vogt:2004hf} $0 \le b \le 3.81$~fm for central,
$4.66 \le b \le 6.01$~fm for semicentral, and $6.59 \le b \le 7.74$~fm 
for peripheral. We will use the notation $R_{CP}$ to refer to the 
central-to-peripheral ratio, while $R_{SP}$ will be used to denote the
semicentral-to-peripheral ratio in the following. 

\subsection{Pseudorapidity Asymmetry}
\label{rapasym}

As the mechanisms for hadron production in d+Au collisions may be different
at forward rapidities (deuteron side) and backward rapidities (gold side),
it is of interest to study ratios of particle yields between a given 
rapidity value and its negative in these collisions. The STAR 
Collaboration has recently measured pseudorapidity 
asymmetries\cite{Abelev:2006pp}, defined as 
\beq
Y_{Asym} = \f{\dd\sigma_{dAu}^{h}}{\dd\eta \ \dd^2p_T}(\mbox{Au-side}) \left/
           \f{\dd\sigma_{dAu}^{h}}{\dd\eta \ \dd^2p_T}(\mbox{d-side}) \right. \,\, ,
\label{yasym}
\eeq
in d+Au collisions for several identified hadron species 
and total charged hadrons in the pseudorapidity intervals 
$|\eta| \le 0.5$ and $0.5 \le |\eta| \le 1.0$. 
Rapidity asymmetries with the
backward/forward ratio above unity for transverse momenta up to $\approx 5$ GeV/c
are observed for charged pion, proton+anti-proton, and total charged hadron production 
in both rapidity regions. We want to see if different nPDFs give significantly different 
rapidity asymmetries for the various hadron species. 

\section{Results}
\label{res}

Before presenting the results from the three different nPDFs, we note that 
while EKS and HKN are similar in the sense that they are global fits to 
experimental data, FGS is relying on Gribov theory for diffractive deep 
inelastic scattering to derive nPDFs. Gluon shadowing is much stronger in
FGS than in EKS and HKN. Because nuclear gluon distributions are poorly 
constrained experimentally, there are significant differences between the
EKS and HKN gluon shadowing. On the other hand, since the Gribov formalism
is not capable to predict valence quark shadowing, FGS uses EKS shadowing for
valence quarks, and all three parameterizations are in agreement for valence
quark shadowing. For sea quarks, EKS and HKN are in good agreement at
$0.01 \lesssim x \lesssim 0.1$. At $x \gtrsim 0.2$ and in the small-$x$ 
region $x \lesssim 0.01$ there are substantial deviations. The FGS predicts 
more nuclear shadowing for sea quarks than EKS and HKN. Due to the $x$-integration,
different nuclear effects [shadowing ($x \lesssim 0.1$, depletion), 
antishadowing ($0.1 \lesssim x \lesssim 0.3$, enhancement), EMC effect ($0.3 \lesssim 
x \lesssim 0.7$, depletion), etc] are superimposed, thus it is difficult to
isolate these effects.

Since pQCD calculations are generally not reliable at low $p_T$, we do not wish 
to push our calculations below $p_T = 1.5$ GeV/c. With our scale choice, this 
corresponds to a minimum $Q^2$ ($Q^{2}_{f}$) of $2.25$ GeV$^2$, while  
EKS, FGS, and HKN give $2.25, 4.0$, and $1.0$ GeV$^2$ for minimum $Q^2$, 
respectively. The minimum $Q^{2}_{f}$ for the AKK fragmentation 
functions is given as $2.0$ GeV$^2$.

\subsection{Nuclear Modification Factors}
\label{rdau}

We have calculated the minimum bias $R_{dAu}$ for total charged hadron production
at the BRAHMS pseudorapidities with the three nuclear parton distribution functions 
considered in this study. In the case of the FGS nPDFs we use the strong
gluon shadowing for the gold nucleus. 
We have only calculated the $p_{T}$ distributions for final-state total charged
hadrons in the present study, since we are mainly interested in the performance of
the different nPDFs. More detail for different hadron species and fractional 
contributions from quarks, antiquarks, and gluons for EKS and FGS nPDFS can be 
found in \cite{Vogt:2004hf}. Here we limit our discussion of nuclear modification
factors to the effect of the different nPDFs. We make two general observations:
\begin{enumerate}
\item[i)] 
At midrapidity (small $\eta$), processes initiated by both gluons and quarks 
are important. At forward rapidities (large $\eta$), $x_{b}$, the parton
momentum fraction in gold becomes small and gluon-initiated processes become
dominant. 
\medskip
\item[ii)] 
In both, the data and the calculations, $R_{dAu}$ decreases systematically with 
increasing $\eta$. This reflects the increasing role of shadowing since
smaller values of $x_{b}$ are probed at forward rapidities.
\end{enumerate} 

\begin{figure*}[htb]
\begin{center} 
\includegraphics[width=16.5cm, height=18.5cm, angle=270]{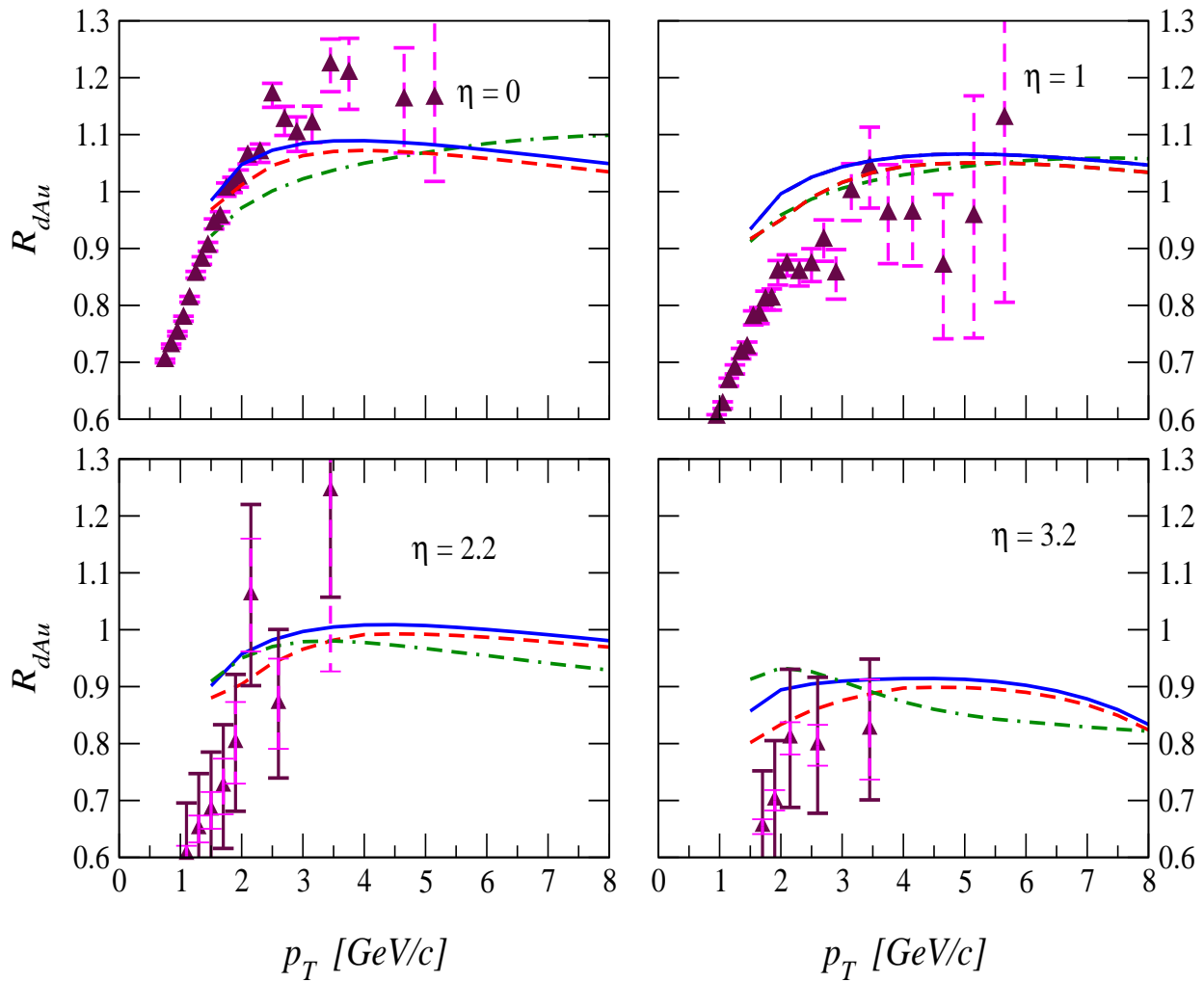}
\end{center}
\caption[...]{(Color Online) Minimum bias nuclear modification factor $R_{dAu}$ 
for total charged hadron production at different pseudorapidities, 
using EKS (solid), FGS (dashed), and HKN (dot-dashed) nPDFs. Solid triangles denote
BRAHMS data. In the top panels the data are for average charged hadrons, while 
for the bottom panels only negative hadrons are measured.
The solid error bars are systematic errors, while dashed
error bars represent statistical errors. For clarity, only statistical errors are 
displayed in the two upper panels, where the systematic errors are rather large.}
\label{fig:rdaum0}
\end{figure*}

The results of our calculations are displayed in Fig.~\ref{fig:rdaum0}, 
together with the experimental data. 
At $\eta = 0$ and $p_{T} = 1.5$ GeV/c, typical $x_{min}$ values are significantly below 0.1. 
Thus there is substantial contribution from the shadowing region,
and all three nPDFs predict $R_{dAu} < 1$, with the HKN being the lowest.
Due to their steeper rise of $R_{dAu}$ with $p_{T}$, the EKS and FGS nPDFs 
appear to describe the data better at around $p_{T} = 3$ GeV/c than the HKN nPDFs. 
At $p_{T} = 4$ GeV/c, $x_{bmin}$ is above 0.5 for some values of $x_{a}$.
There is thus more contribution from the antishadowing region, 
and consequently 
$R_{dAu} > 1$, with the HKN still being the lowest.  At $p_{T} = 8$ GeV/c,
we have major contributions from both antishadowing and the EMC effect. Thus, 
while $R_{dAu}$ is still $> 1$, the trend is towards $1$ for both EKS and FGS. 
The HKN parameterization predicts a higher value at around 1.1. 

The behavior is similar at $\eta = 1$. At $p_{T} = 1.5$ GeV/c both $x_{min}$
values are small, and thus $R_{dAu} < 1$. Around
this $p_{T}$ the HKN is identical with the FGS. At $p_{T} = 4$~GeV/c, 
$x_{bmin}$ ranges higher, with $R_{dAu} > 1$. The HKN is 
slightly below both EKS and FGS in this region. At $p_{T} = 8$ GeV/c, 
$R_{dAu} > 1$ with substantial contributions from both antishadowing and the 
EMC effect. The HKN is practically the same as EKS and slightly higher than FGS. 
All three are in good agreement with data except at low $p_T$, where the data are 
more suppressed than the calculated results.

The effect of increasing $\eta$ is already apparent at $\eta = 2.2$, where both 
FGS and HKN are below unity for all $p_{T}$ considered. The EKS is practically
$1$ at mid-$p_T$ and falls below $1$ at high $p_T$.  
The major contribution is from shadowing with the resultant $R_{dAu} < 1$. 
Note that at low $p_T$ the HKN is similar to the EKS, while 
at higher $p_T$ it is lower than both EKS and FGS. The agreement with data 
is reasonable.

At $\eta = 3.2$ 
the dominant contribution is again
from shadowing with $R_{dAu} < 1$ for all $p_T$. At low $p_T$ the FGS
describes the data best while at high $p_T$ it is similar to the EKS. This 
may be due to the stronger gluon shadowing in the FGS. The HKN is less 
suppressed at low $p_T$ than both EKS and FGS, while at higher $p_T$ it is 
more suppressed than the others..

\subsection{Central-to-Peripheral Ratios}
\label{rcp}

\begin{figure*}[htb]
\begin{center} 
\includegraphics[width=16.5cm, height=18.5cm, angle=270]{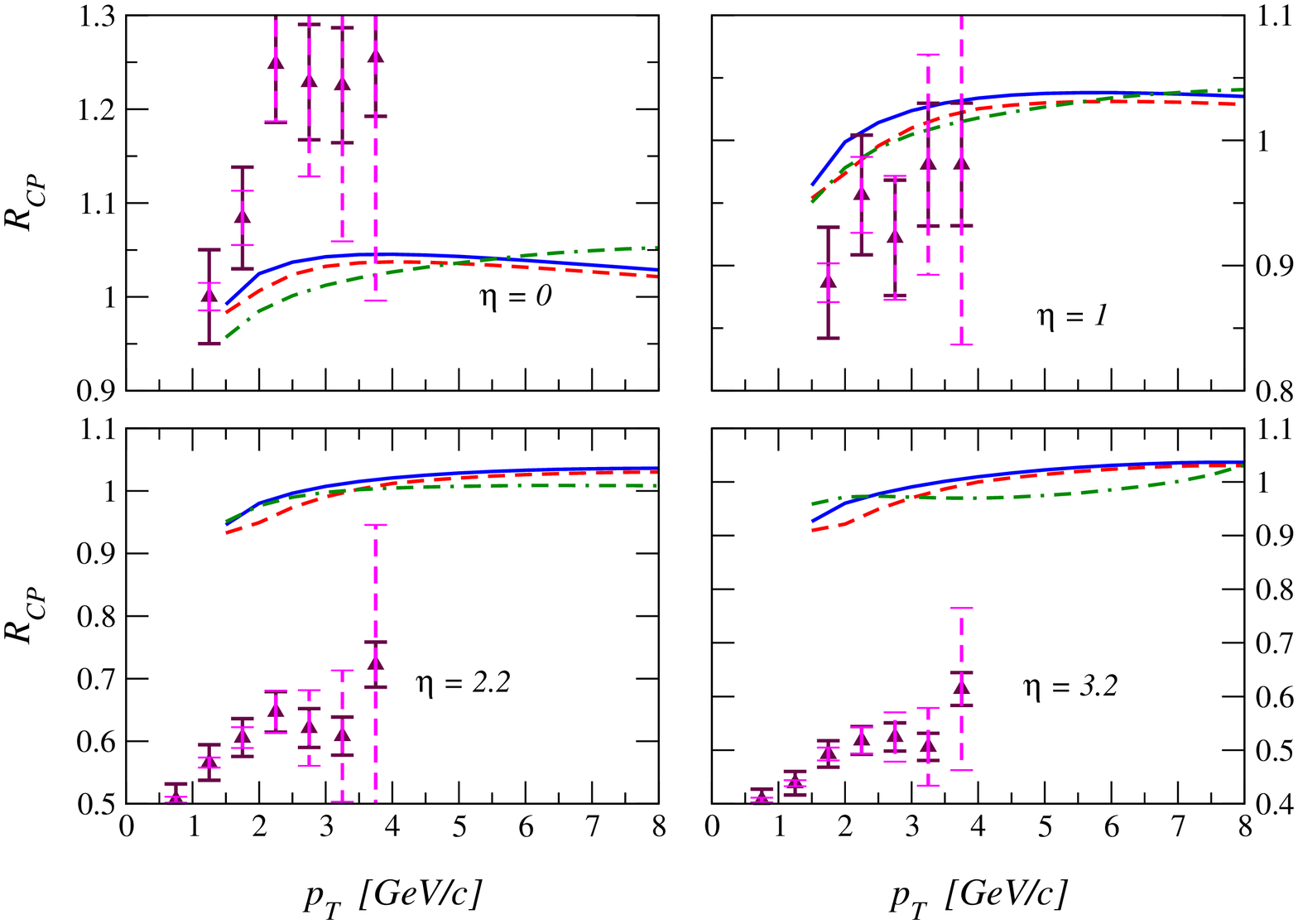}
\end{center}
\caption[...]{(Color Online) 
Central-to-peripheral ratio $R_{CP}$ for total charged hadron production
at different pseudorapidities using EKS (solid), FGS (dashed), and 
HKN (dot-dashed) nPDFs. Solid triangles denote
BRAHMS data. The solid error bars are systematic errors, while dashed
error bars represent statistical errors.}
\label{fig:rcp3}
\end{figure*}

\begin{figure*}[htb]
\begin{center} 
\includegraphics[width=16.5cm, height=18.5cm, angle=270]{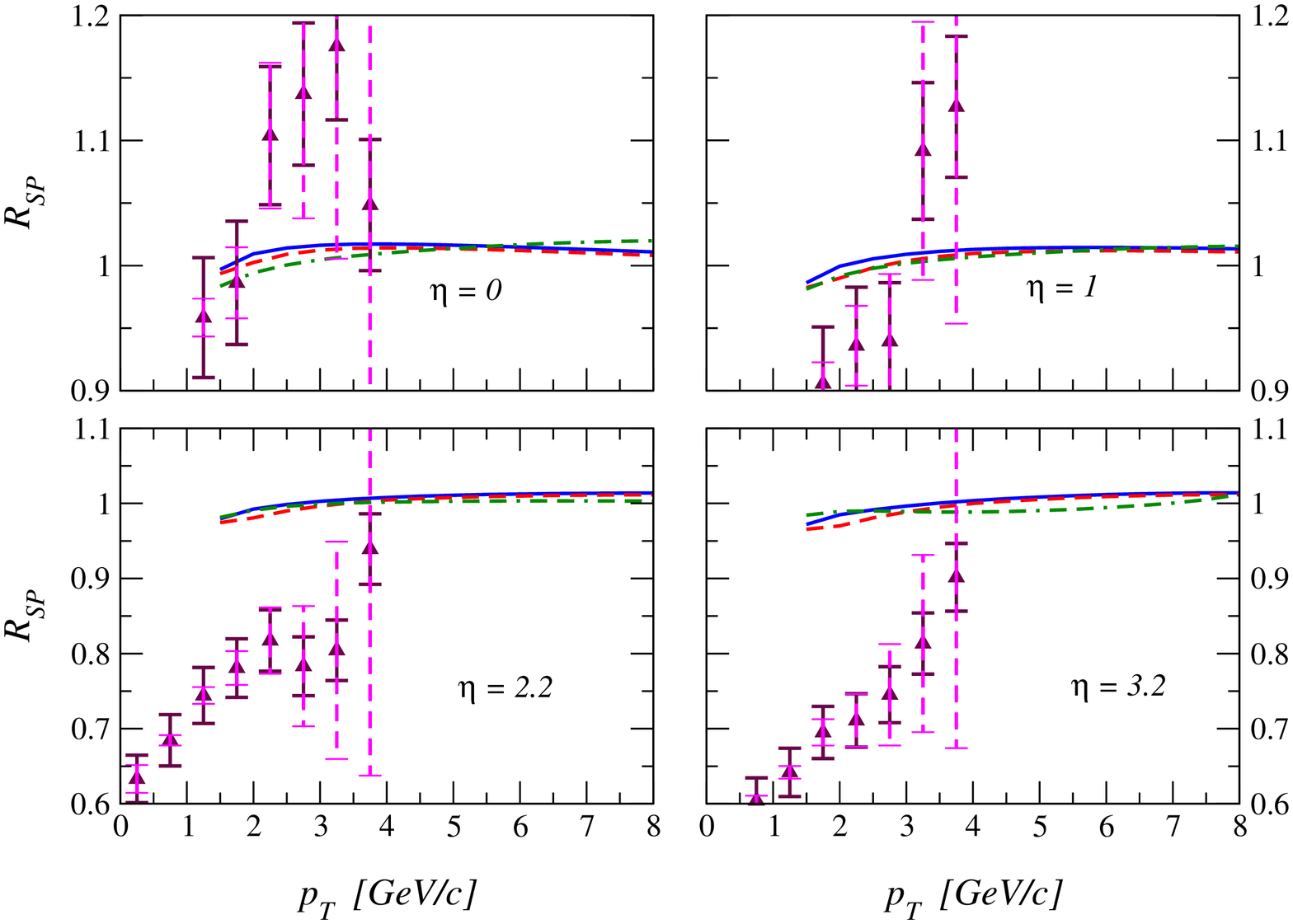}
\end{center}
\caption[...]{(Color Online) 
Semicentral-to-peripheral ratio $R_{SP}$ for total charged hadron production
at different pseudorapidities using EKS (solid), FGS (dashed), and 
HKN (dot-dashed) nPDFs. Solid triangles denote
BRAHMS data. The solid error bars are systematic errors, while dashed
error bars represent statistical errors.}
\label{fig:rsp3}
\end{figure*}

The results of our central-to-peripheral ratio ($R_{CP}$) calculation 
are displayed in Fig.~\ref{fig:rcp3}, together with the BRAHMS data.  At 
$\eta = 0$, where the data indicate an $R_{CP} > 1$ at 
$2 \lesssim p_T \lesssim 4$~GeV/c, the calculated results are below the
data, while at both $\eta = 2.2$ and $\eta = 3.2$ the data show significant
suppression, with the calculations giving $R_{CP}$ close to unity. At $\eta = 1$ 
the calculation can be said to be in reasonable agreement with the data. 
Although the calculation exhibits the trend toward increasing suppression in
the data as $\eta$ increases, the calculated variation with $\eta$ is much 
smaller than the one shown by the data. This shortcoming becomes increasingly
evident at forward rapidities. 
Fig.~\ref{fig:rsp3} shows the calculated semicentral-to-peripheral ratio,
$R_{SP}$,  using the three nPDFs, together
with the BRAHMS data. The situation here mirrors that 
of the central-to-peripheral ratios. The results underpredict the data at 
$\eta = 0$, but overpredict at forward rapidities. The degree 
of suppression at forward rapidities is not as severe as in the 
central-to-peripheral ratios. In fact, due to the large error bars, there 
is reasonable agreement with data at $p_{T}$ around 4~GeV/c. 

Both $R_{CP}$ and $R_{SP}$ are geometry-dependent, thus the 
assumed spatial dependence of shadowing is important. We have checked that  
shadowing proportional to the local density gives worse agreement 
with data than shadowing proportional to the thickness function.
A variation of the thickness function dependence where we used higher powers 
gives better agreement at $\eta = 0$ but still overpredicts significantly at other 
rapidities. With these shadowing parameterizations, a more radical spatial 
dependence is needed to describe the data for both ratios. Another factor that 
may be responsible 
is a too-weak $x$-dependence of the available shadowing parameterizations.

\subsection{Pseudorapidity Asymmetry}
\label{raprats}

In d+Au collisions at small rapidities, particle production may include 
contributions from gold-side partons that may have been modified by nuclear 
effects and from deuteron-side partons that have experienced multiple 
scatterings while traversing the gold nucleus \cite{Abelev:2006pp}. It should 
be kept in mind that the latter effect is not included in the present calculations.

We have calculated $Y_{asym}$ for charged pions 
($\pi^+ + \pi^-$), charged kaons, protons+antiprotons ($p + {\bar p}$),
and total charged hadrons ($h^+ + h^-$). Below 2 GeV/c, data are available 
for total charged hadrons only. Above 2 GeV/c, separated data
exist for $\pi^+ + \pi^-$ and for $p + {\bar p}$. A benefit of
using a ratio is that the systematic errors largely cancel and are 
$\lesssim 5\%$ for both pions and protons, $< 3\%$ for charged hadrons 
\cite{Abelev:2006pp}. For $p_{T} \lesssim 4.2$ GeV/c the errors are dominantly
systematic and thus the statistical errors are not displayed. At higher
$p_{T}$ statistical errors tend to become dominant.
Since the asymmetry is a ratio of the yields in two different rapidity
intervals, the respective $x_b$-distributions are mostly responsible for the 
global trends observed in the calculations.

\begin{figure*}[htb]
\begin{center} 
\includegraphics[width=16.5cm, height=18.5cm, angle=270]{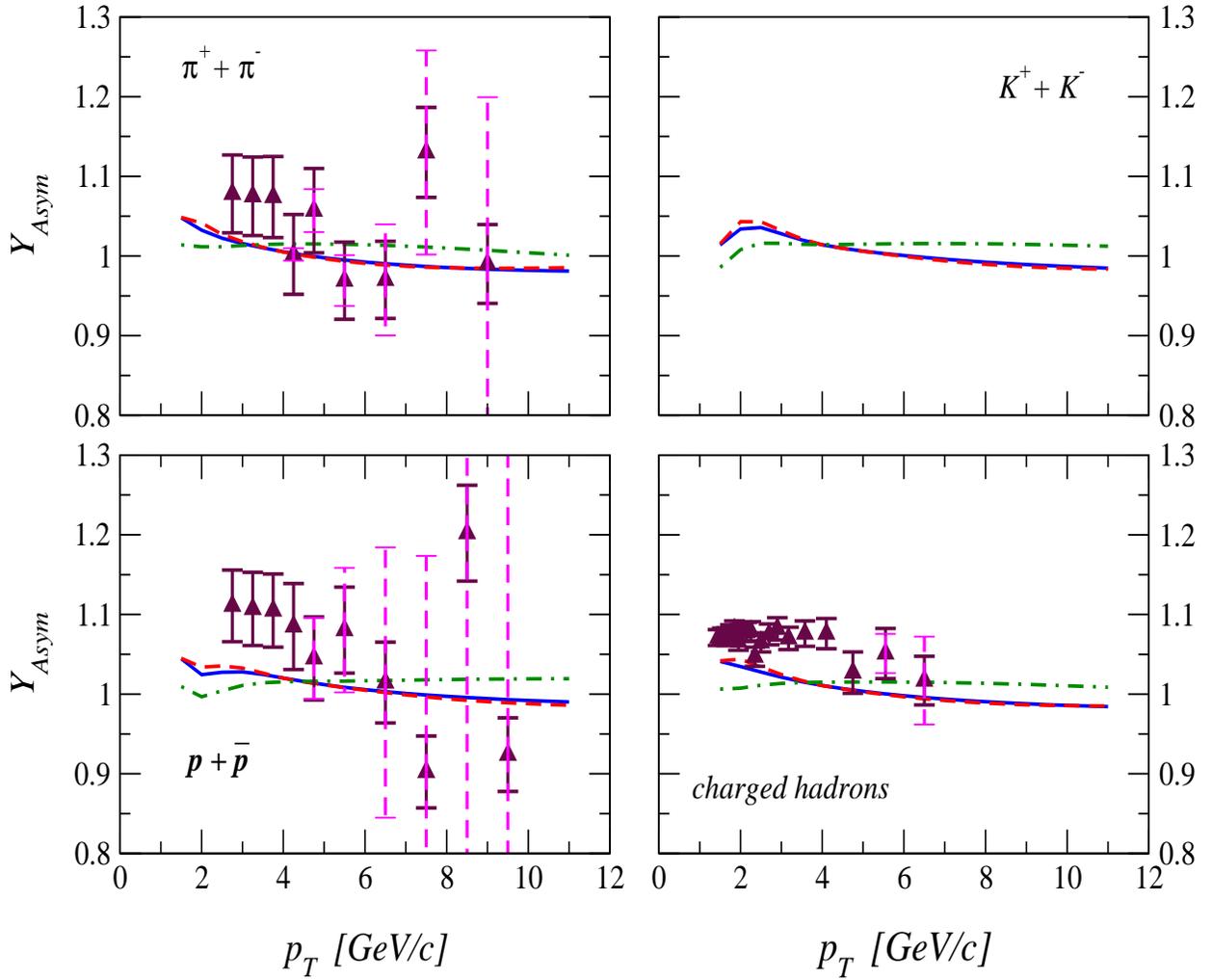}
\end{center}
\caption[...]{(Color Online) Pseudorapidity asymmetry,
$Y_{asym}$ for different hadrons at $|\eta| \le 0.5$. The solid line represents the
EKS nPDFs while the dashed line is obtained with the FGS nPDFs. The dot-dashed line 
corresponds to the HKN nPDFs, and filled triangles denote 
the STAR data. The solid error bars are systematic errors, while dashed error bars 
represent statistical errors.}
\label{fig:asy005}
\end{figure*}

Fig.~\ref{fig:asy005} shows the result of our pseudorapidity asymmetry ratio
calculations for the interval $|\eta| \le 0.5$, together with
the STAR data\cite{Abelev:2006pp}. We plot the calculated 
results with the EKS (solid line), FGS (dashed), and HKN (dot-dashed)
nuclear parton distributions. 
Around $\eta = 0$ one expects the degree of asymmetry to be small. This is 
borne out by both data and calculation. At very low $p_T$, there is more 
contribution from the shadowing region for the positive rapidity
(deuteron-side) than the negative rapidity (gold-side). Thus the ratio is
expected to be above unity. For $p_T > 3$ GeV/c, the $x_b$ distributions are 
similar, with less contribution from the shadowing region as $p_T$
increases. Thus the asymmetry is not far from unity.
The EKS and FGS nPDFs give similar results 
for all hadronic species. The HKN nPDFs yield a different curvature 
and pseudorapidity asymmetries that remain above unity for a wider range 
of transverse momenta. For the total charged 
hadrons ratio, the calculation is below the data for $p_{T} \lesssim 5$ GeV/c. 
At the present level of combined experimental and theoretical uncertainties, 
we can claim that $Y_{asym}$ is
reasonably reproduced for $\pi^+ + \pi^-$ and $p + {\bar p}$, in particular at
large transverse momenta. The EKS and FGS predictions 
for charged kaons have a small maximum at $p_T \approx 2.2$ GeV/c.

\begin{figure*}[htb]
\begin{center} 
\includegraphics[width=16.5cm, height=18.5cm, angle=270]{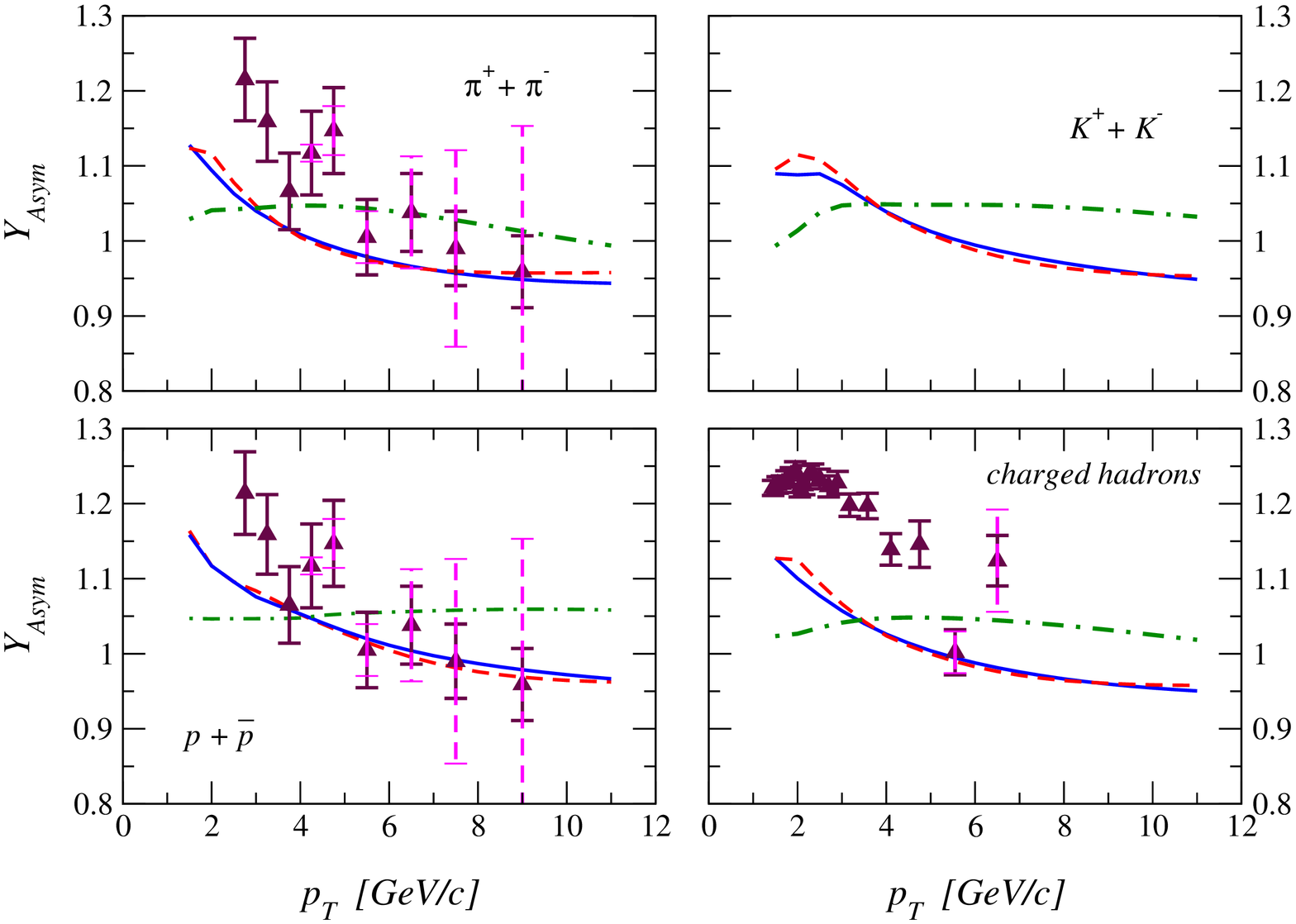}
\end{center}
\caption[...]{(Color Online) Pseudorapidity asymmetry,
$Y_{asym}$ for different hadrons at $0.5 \le |\eta| \le 1.0$. The solid line 
uses EKS nPDFs, while the dashed line is obtained with the FGS nPDFs. The dot-dashed 
line corresponds to the HKN nPDFs, and filled triangles denote 
the STAR data. The solid error bars are systematic errors, while dashed error bars 
represent statistical errors.}
\label{fig:asy051}
\end{figure*}

Figure~\ref{fig:asy051} shows the result of our pseudorapidity asymmetry ratio
calculations for the interval $0.5 \le |\eta| \le 1.0$. At low $p_T$, the 
situation is similar to $|\eta| \le 0.5$. Thus the asymmetry is above unity. 
However, at $p_T > 3$ GeV/c, the $x_b$ distributions start to become significantly 
different. While there are still some contributions from the shadowing region for the
positive rapidity even up to the highest $p_T$, the negative rapidity yield has no 
shadowing contribution for $p_T \gtrsim 8$ GeV/c. Thus one expects the
asymmetry to be more substantial than for $|\eta| \le 0.5$. This is borne out 
by the calculations.
As in $|\eta| \le 0.5$,
EKS and FGS give very similar results. The HKN nPDFs also behave 
similarly to what was seen at $|\eta| \le 0.5$, the difference between EKS and FGS
on the one side, and HKN on the other, becoming more pronounced.
All calculations 
underpredict the data for $h^+ + h^-$. For $\pi^+ + \pi^-$ and $p + {\bar p}$, 
the calculation agrees with the data within error for large transverse momenta.
It will be interesting to see the data for charged kaons, when they become 
available\cite{Xu2007}.
\section{Conclusion}
\label{concl}

We have calculated nuclear modification factors $R_{dAu}$
in deuteron-gold collisions at $\sqrt{s} = 200$~GeV
using three nuclear shadowing parameterizations: 
the EKS, FGS, and HKN nuclear parton distribution functions. We have also 
calculated a related modification factor, the central-to-peripheral ratio, 
$R_{CP}$, using the impact parameter representation of the BRAHMS centrality
classes. We have used the AKK fragmentation functions throughout in our calculations.

All three nPDFs give similar results for the minimum bias $R_{dAu}$ at forward 
rapidities. For not too large rapidities and relatively low  $p_{T}$, the HKN 
values are smaller than those of the EKS and FGS.  The calculated 
results for all three nPDFs approximately agree with the data.

The $R_{CP}$ results do not describe the data well overall. At $|\eta| \le 0.2$ 
the calculated results underestimate the data, while at the most forward rapidities 
the data are much below the calculated results. Only at $\eta = 1$ 
can we claim approximate agreement with data, but this appears to be coincidental,
judging from the $\eta$-dependence of the data and the calculated results.
This situation may be due to poor knowledge of the spatial dependence of
nuclear shadowing. It may also signify a rapidity dependence of shadowing. 

For all hadronic species the calculated pseudorapidity asymmetry underestimates 
the data for $p_{T} \lesssim 5$ GeV/c. At higher $p_{T}$, EKS and FGS results 
tend to fall lower than unity, while HKN yields values above one. Present
uncertainties in the experimental data do not allow a distinction between these 
scenarios. We expect further data up to higher transverse momenta and with smaller
uncertainties in the near future for all the observables discussed here.

\section{Acknowledgments}
\label{ack}

This work was partly supported by the U.S. Department of Energy under
grant U.S. DE-FG02-86ER40251. We thank V. Guzey for stimulating discussions and 
B.~ Mohanty for information on experimental errors. 

%

\end{document}